\documentclass[preprint,5p]{elsarticle}
\pdfoutput=1

\usepackage{graphicx}
\usepackage{dcolumn}
\usepackage{bm}
\usepackage{siunitx}
\usepackage{color}
\usepackage{amsmath,amssymb,amsthm,color}
\usepackage{amsbsy,amssymb,latexsym,amsfonts,amsmath}

\allowdisplaybreaks

\newcommand{\Slash}[1]{{\ooalign{\hfil#1\hfil\crcr\raise.167ex\hbox{/}}}}

\begin{document}


\title{Non-trivial Center Dominance in High Temperature QCD }

\author{K.-I. Ishikawa}
\address{Graduate School of Science, Hiroshima University,Higashi-Hiroshima, Hiroshima 739-8526, Japan}

\author{Y. Iwasaki}
\address{Center for Computational Sciences, University of Tsukuba,Tsukuba, Ibaraki 305-8577, Japan}

\author{Yu Nakayama}
\address{ Walter Burke Institute for Theoretical Physics,California Institute of Technology,  Pasadena, CA 91125, USA;\\
Kavli Institute for the Physics and Mathematics of the Universe (WPI),
Todai Institutes for Advanced Study, Kashiwa, Chiba 277-8583, Japan}

\author{T. Yoshie}
\address{Center for Computational Sciences, University of Tsukuba,Tsukuba, Ibaraki 305-8577, Japan}

\date{\today}

\begin{abstract}
	We investigate the properties of quarks and gluons above the chiral phase transition temperature $T_c,$ using the RG improved gauge action and the Wilson quark action with two degenerate quarks mainly on a $32^3\times 16$ lattice.
	In the one-loop perturbation theory, the thermal ensemble is dominated by the gauge configurations with effectively $Z(3)$ center twisted boundary conditions, making the thermal expectation value of the spatial Polyakov loop take a  non-trivial $Z(3)$ center.
	This is in agreement with our lattice simulation of high temperature QCD.
	We further observe that the temporal propagator of massless quarks at extremely high temperature $\beta=100.0 \,  (T \simeq10^{58} T_c)$ remarkably agrees  with the temporal propagator of free quarks with the $Z(3)$ twisted boundary condition for $t/L_t \geq 0.2$, but differs from that with the $Z(3)$ trivial boundary condition. As we increase the mass of quarks $m_q$, we find that the thermal ensemble continues to be dominated by the $Z(3)$ twisted gauge field configurations as long as $m_q \le 3.0 \,T$ and above that the $Z(3)$ trivial configurations come in.
The transition is similar to what we found in the departure from the conformal region in the zero-temperature  many-flavor conformal QCD on a finite lattice by increasing the mass of quarks.
\end{abstract}

\maketitle

\section{introduction}

The properties of quarks and gluons at high temperature are key ingredients for understanding the
evolution of the Universe and the heavy ion collision experiment.
Lattice QCD is the most reliable formulation of QCD for the investigation of  non-perturbative properties of
quarks and gluons, and  we have developed various methods to clarify them from the early stage of lattice gauge theories
\cite{pioneer}.
Although many interesting and useful results have been obtained,
still there remain unsolved problems \cite{review}.
In particular, it is a fundamental issue to understand what kind of state the gluons and quarks take at high temperature\cite{eq of state}.

In this article we give a new perspective on the properties of
quark-gluon state above the chiral phase transition temperature $T_c.$
Our analysis is based on the numerical simulations of spatial Polyakov loops and temporal propagator in the pseudo-scalar (PS) channel at finite temperature in comparison with those of the one-loop perturbative computation with the fixed boundary conditions. Our result suggests that for light quarks, the thermal path integral is dominated by the gauge cofiguration with the non-trivial center.

Some related results of our study on the high temperature QCD  have been presented in Refs.\cite{Ishikawa:2013iia},\cite{PR89}.
The theoretical argument presented there is applicable to any aspect ratio $r=N_s/N_t$. The numerical simulations there were done on the $16^3\times 64$ lattice  with $r=1/4$ for a technical reason to obtain the large $t$ behavior of PS propagators. In applications to the high temperature QCD, however, the larger aspect ratio is preferable with thermodynamic limit in mind. 
In this article, therefore, we extend the analysis to larger aspect ratio and observe the similar behavior.


The organization of the paper is as follows. After describing our setup in section II,  we discuss the phase diagram with respect to mass and bare gauge coupling  in section III. In section IV, we discus  the properties of the phase diagram at weak coupling based  on the numerical simulations
in comparison with 
the one-loop perturbative calculations.
In section V we investigate the phase structure
for a wide range of the coupling
along the massless quark line  toward the chiral transition point.
In section IV and V, we also investigate the case of the massive quark,
and present the evidence for the transition.
We conclude the paper with further discussions in section VI.
In Appendix A, we report the one-loop calculation of the internal energy on a finite lattice.

\section{Setup}
We investigate SU(3) gauge theory at high temperature
 with small $N_f$ ($2 \le N_f \le 6$) fermions in the fundamental representation as a model of QCD, where the chiral phase transition occurs at some critical temperature $T_c$.
Our general argument that follows can be applied to any number of flavors ($2 \le N_f \le 6$)
with any formulation of  gauge theories on the lattice. 
For numerical simulations in this article we take $N_f=2$ (degenerate two quarks) and employ the Wilson quark action and the RG improved gauge action\cite{RG-improved}
 on the Euclidean lattice of the size $N_x=N_y=N_z=N_s$ and $N_t,$ 
with the lattice spacing $a$. We impose an anti-periodic boundary condition in the time direction for fermions and periodic boundary conditions otherwise.

In order to take the continuum limit, we have to take the limit $a \rightarrow 0$ keeping
$N_s\, a=L_s$ and $N_t \, a=L_t$ constant.
We call $r=L_s/L_t$ an aspect ratio. 
In order to obtain physical quantities at temperature $T$,
we have to take the thermodynamic limit $L_s\rightarrow \infty$  
keeping $N_t\, a =1/T$ and  QCD scale $\Lambda_{\mathrm{QCD}}$ fixed. 
When the space is compact, $N_s\, a =L_s$ is also fixed finite.

When $N_s$ and $N_t$  are finite, the formulation of finite temperature QCD on a lattice is equivalent to a Euclidean path integral defined on a discrete three dimensional degrees of freedom with a transfer matrix for a discrete time. We may calculate thermal quantities for the quantum system in terms of the terminology of zero-temperature field theories. Therefore, our analysis of the lattice data naturally shows the similarity to the ones studied in our earlier papers \cite{PR87}\cite{PR89}.



Given a fixed lattice, the theory is defined by two parameters; the bare coupling constant $g_0$ and the bare degenerate quark mass $m_0$ at ultraviolet (UV) cutoff.
We also use, instead of $g_0$ and $m_0$, 
$\beta={6}/{g_0^2}$
and 
$K= 1/2(m_0a+4)$.  

As for observables, together with the plaquette and the Polyakov loop in each space-time direction, we measure several hadronic quantities.
One of the most important observables we will study is the $t$ dependence of the thermal propagator of the local meson operator in the $H$ channel:
\begin{equation}
G_H(t) = \sum_{x} \langle \bar{\psi}\gamma_H \psi(x,t) \bar{\psi} \gamma_H \psi(0) \rangle \ ,
\label{propagator}
\end{equation}
where the summation is over all the spatial lattice points.
In this paper, we mostly focus on the pseudo-scalar (PS) channel $H=PS$, and the subscript $H$ is suppressed hereafter.

In order to study the characteristic behavior of the thermal propagator, 
we define the effective mass $m(t)$ through  
\begin{equation}\frac{\cosh(m(t)(t-N_t/2))}{\cosh(m(t)(t+1-N_t/2))}=\frac{G(t)}{G(t+1)}.\label{effective mass}\end{equation}
When boundary effects can be neglected, it reduces to
\begin{equation}m(t) = \ln \frac{G(t)}{G(t+1)}.\label{simple effective mass}\end{equation}
This is the same definition that we use in the zero-temperature lattice QCD.
The notion of effective mass has its direct physical interpretation at zero temperature, but we may regard it as a characteristic of the temporal propagator at finite temperature to compare thermal systems with different parameters.

We define the quark mass $m_q$ as the large $t$ value of $m_q(t)$ obtained through Ward-Takahashi identities by the ratio of thermal propagators:
\begin{align}
m_q &= \lim_{t  \to N_t/2}m_q(t) \cr
  &= \lim_{t  \to N_t/2} \frac {\sum_{x} \langle \nabla_4 A_4(x,t) P(0) \rangle }{ 2\, \sum_{x} \langle P(x,t) P(0) \rangle}
\label{quark mass}
\end{align}
where $P(x,t)$ is the pseudo-scalar density and $A_4(x,t)$ the fourth component of the
local axial vector current, renormalization constants being suppressed.
This is also the same definition that we use in the zero-temperature lattice QCD.
The quark mass $m_q$ thus defined does not depend on whether the system is confining or deconfining, and
depends on only $\beta$ and $K$ up to order $1/N_s$ and $1/N_t$  corrections.


\begin{table}
\caption{Job parameters: $\beta$ and $K$}
\begin{tabular}{lrrrrrrr}
\hline
\hline
$\beta$ \, & 100.0 & 15.0 &  10.0 &  6.0 & 5.0   & 4.0  &3.0\\
$K$ \, & 		0.125 & 0.128  & 0.130 & 0.133 &	   0.135 &0.140    &    0.1435\\
\hline
$T/T_c\, \, $ & $10^{58}$ &$10^{7}$ & $10^{4}$ & 64 & 16 &4 & 1.15\\
\hline
\end{tabular}
\end{table}

We perform the simulations on the $32^3 \times 16$ lattice with the parameters given in Table 1: We take the hopping parameters $K$  in such a way that the quark mass satisfies $|m_q| \leq 0.01.$
Here the rough estimate of the temperatures is based on $\Delta \beta \sim 0.5$ for the scale change of a factor $2$ in the one-loop approximation of the beta function with $N_f=2$.

The algorithms we employ is the blocked HMC algorithm \cite{Hayakawa:2010gm}.
We choose the run-parameters in such a way that the acceptance of the HMC Metropolis test is about $60\%\sim 90\%.$
The statistics are 1,000 MD trajectories for thermalization and $1000\sim6000$ MD trajectories for the measurement.
We estimate the errors by the jack-knife method
with a bin size corresponding to 100 HMC trajectories.

\section{Phase structure}
We define continuous gauge theories by the continuum limit of lattice gauge theories.
Therefore, in order to investigate properties of the high temperature QCD in the continuum limit,
it is vital to clarify the phase structure of lattice QCD with $N_s$ and $N_t$ fixed, thereby clarify the existence of fixed points and symmetries.  
We plot the two dimensional diagram in terms of $\beta$ and $K$ for a fixed value of $N_s$ and $N_t$ in Fig.1.

First of all, the point $\beta=\infty$ and $m_q=0$ is the UV fixed point and we restrict ourselves to the continuum limit toward this UV fixed point in this article. 
When $N_f=2\sim 6$, the massless quark line starting from the UV fixed point ($\beta=\infty$) runs through to $\beta=0$, without hitting a bulk transition.
On a finite lattice there is a chiral phase transition at some $\beta^*$. At $\beta \le \beta^*$ the system is in the confining phase. 
On the other hand, at $\beta \ge \beta^*$ the system is in the deconfining phase.

In Refs.\cite{iwa96},\cite{iwa97}
we estimated the chiral phase transition of $N_f=2$ QCD with the same action on an $8^3\times 4$ lattice at $\beta\simeq 1.4.$ 
Similarly, applying the ``on-Kc method" \cite{iwa96} (monitoring the number of iterations for the quark matrix inversion along the Kc line), we estimate that the transition points are  $\beta \sim 2.4$ and  $K \sim 0.154$ on the $16^3\times 8$ lattice and $\beta \sim 2.9$ and $K \sim 0.1445$ on the $32^3\times 16$ lattice, respectively.

In addition to these two phases, we claim that there is another region in which the gauge configuration with the non-trivial center dominates for the small quark mass, $m_q \le c\, \Lambda_{\mathrm{IR}},$ 
$(\Lambda_{\mathrm{IR}}$  an IR cutoff; $c$ is a constant of $O(1 \sim 10 $)\cite{PR87}\cite{PR89}), as schematically shown in Fig.1.

\begin{figure}[thb]
	\includegraphics [width=7.5cm]{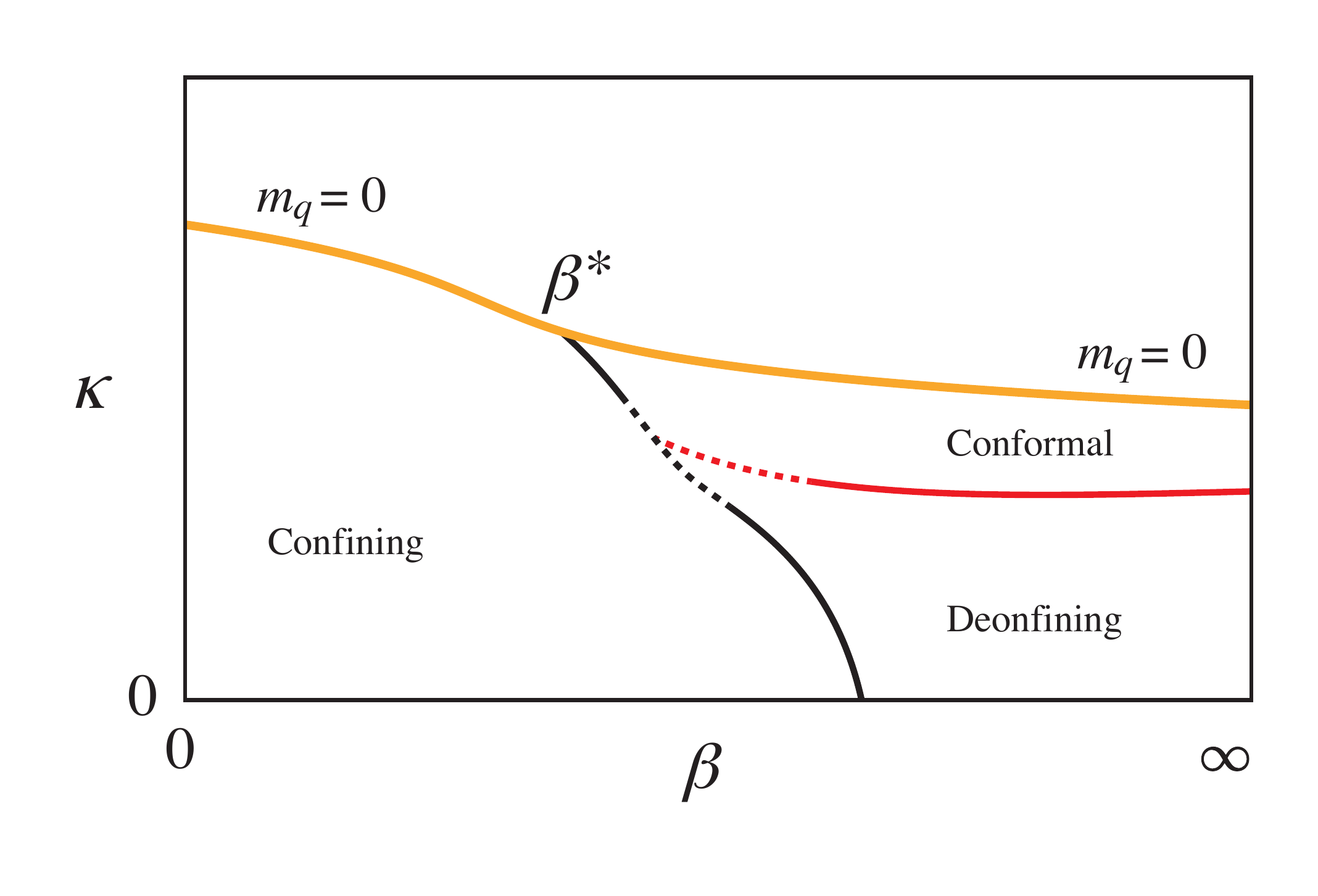}
	\caption{ Phase diagram on a finite lattice : $2 \le N_f \le 6$ ;   on the massless quark line there is a chiral phase transition point $\beta^{*}$. Below the critical point the massless line is in the confining region.}
	\label{phase diagram finite lattice}
\end{figure}


The situation is very similar to the study of the many-flavor conformal QCD put on a finite lattice \cite{PR87}.
In the many-flavor conformal QCD at zero temperature, if the IR cutoff are zero and
when quark masses are tiny,  the RG trajectory stays close to the critical line, approaching the IR fixed point and finally passes away from the IR fixed point to infinity. Therefore the IR behavior is governed by the ``confining region''. 

When the cutoff $\Lambda_{\mathrm{IR}}$ is finite, the RG flow from UV to IR does stop evolving at the scale $\Lambda_{\mathrm{IR}}$.
When the typical mass scale (e.g. that of a meson) $m_H$ is smaller than $\Lambda_{\mathrm{IR}}$ (it means $m_q \leq c \Lambda_{\mathrm{IR}}$), it is still in the ``conformal region'' because the IR cut-off rather than the mass scale governs the IR behavior.
On the other hand, when $m_H$ is larger than $\Lambda_{\mathrm{IR}}$, the flow passes  away from the IR fixed point to infinity  with relevant variables integrated out, thus being in the  ``confining region''.

One may alternatively view that the conformal region is the parameter region in which the dominant scale of the system is given by an IR cut-off (rather than mass scales). In such situations, the effects by an IR cut-off cannot be ignored even when the lattice size is large.
We have introduced the terminology ``regions" in contrast to the phases because strictly speaking there are no order parameters in the thermodynamic limit to distinguish them. However, we claim that  the boundary between these regions accompanies a sharp transition in physical observables. This has been confirmed in our numerical simulations \cite{PR87}\cite{PR89}.

The situations in the high temperature QCD ($2 \leq N_f \leq 6$) is
very similar \cite{PR89}\cite{Ishikawa:2013iia} even though there is
another scale $\Lambda_{\mathrm{QCD}}$. As long as we are in the
deconfining phase with tiny masses for quarks, the dominant scale is
the IR cut-off given by the temperature $T$ rather than
$\Lambda_{\mathrm{QCD}}$. Similarly to the many-flavor conformal QCD case with  a
spatial IR cut-off, there is no response for the RG transformation
below the scale of the temperature $\Lambda_{\mathrm{IR}} = T$.
In the following we are going to pursue this similarity in a more precise way with the lattice numerical simulations.




%


\begin{figure*}[htb]
	\includegraphics [width=7.7cm]{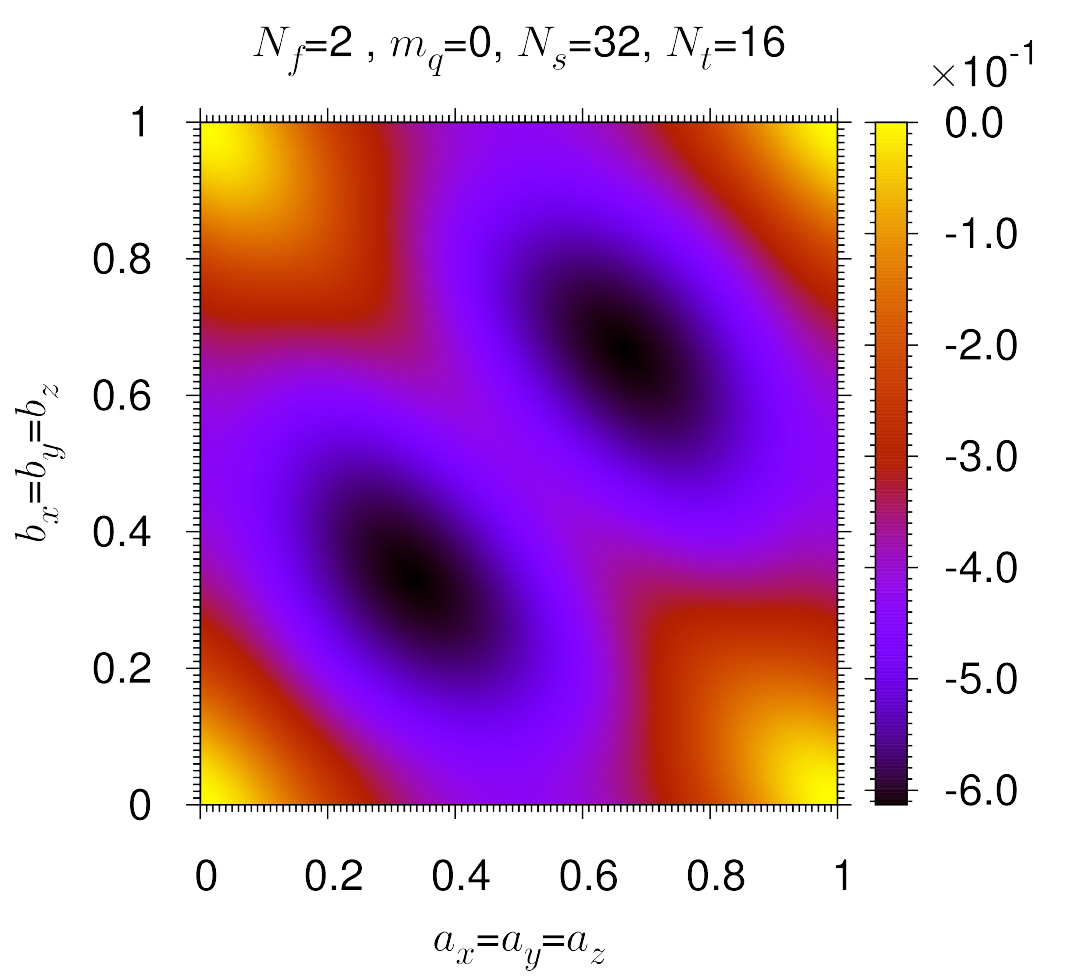}
	\hspace{1cm}
	\includegraphics [width=7.7cm]{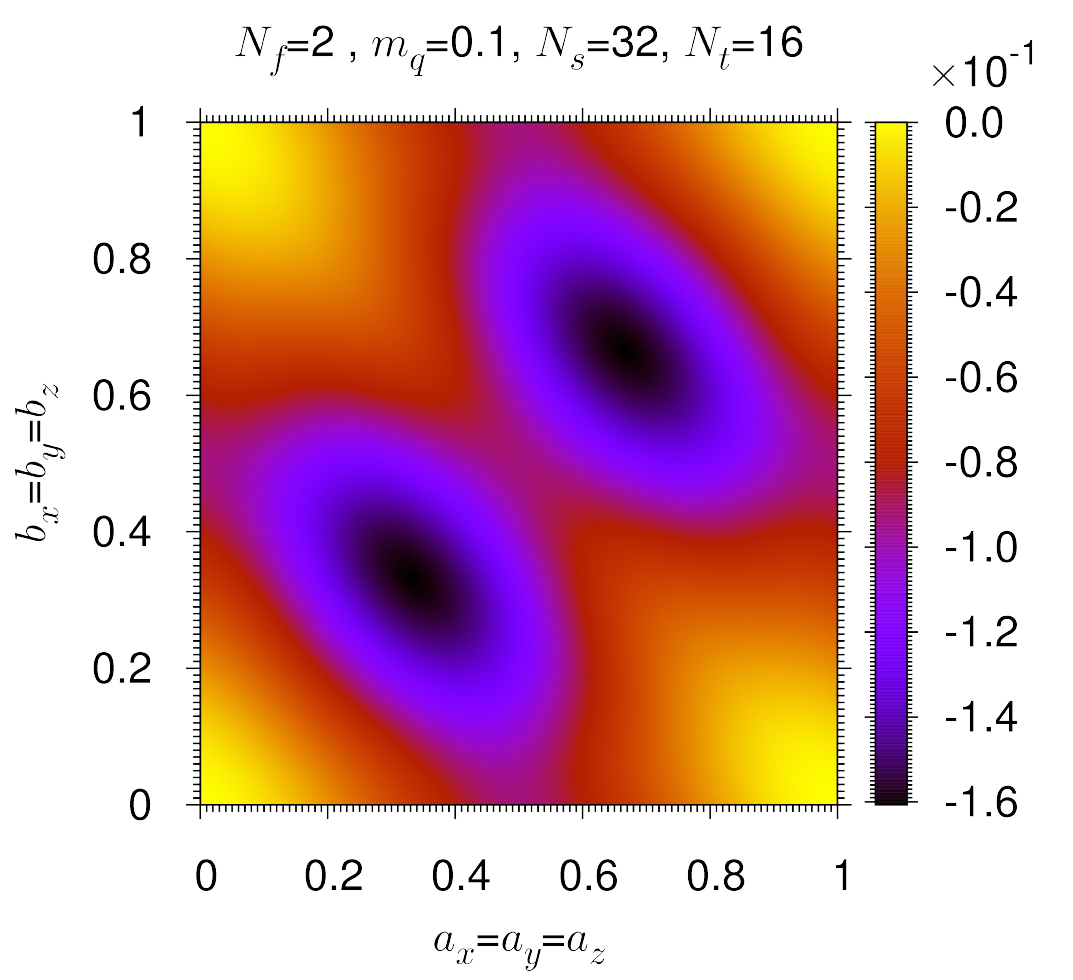}
		\caption{(color online)  The effective potential $V_{\mathrm{eff}}(a, b)$ on $32^3\times 16$ lattice in terms of $a$ and $b$: $m=0.0$ (left) and $m=0.1$ (right).}
	\label{effective potential 1}
\end{figure*}

\begin{figure*}[htb]
	\includegraphics [width=7.5cm]{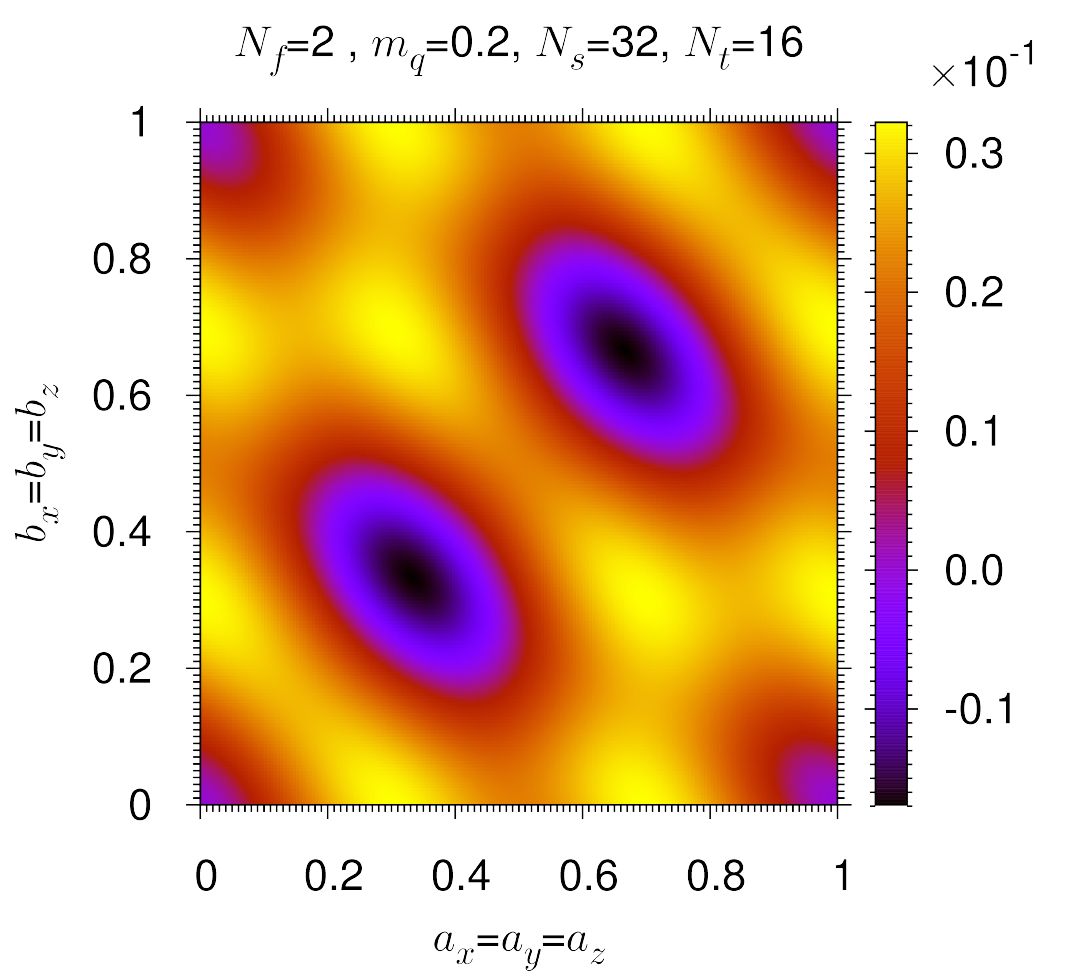}
	\hspace{1cm}
	\includegraphics [width=7.5cm]{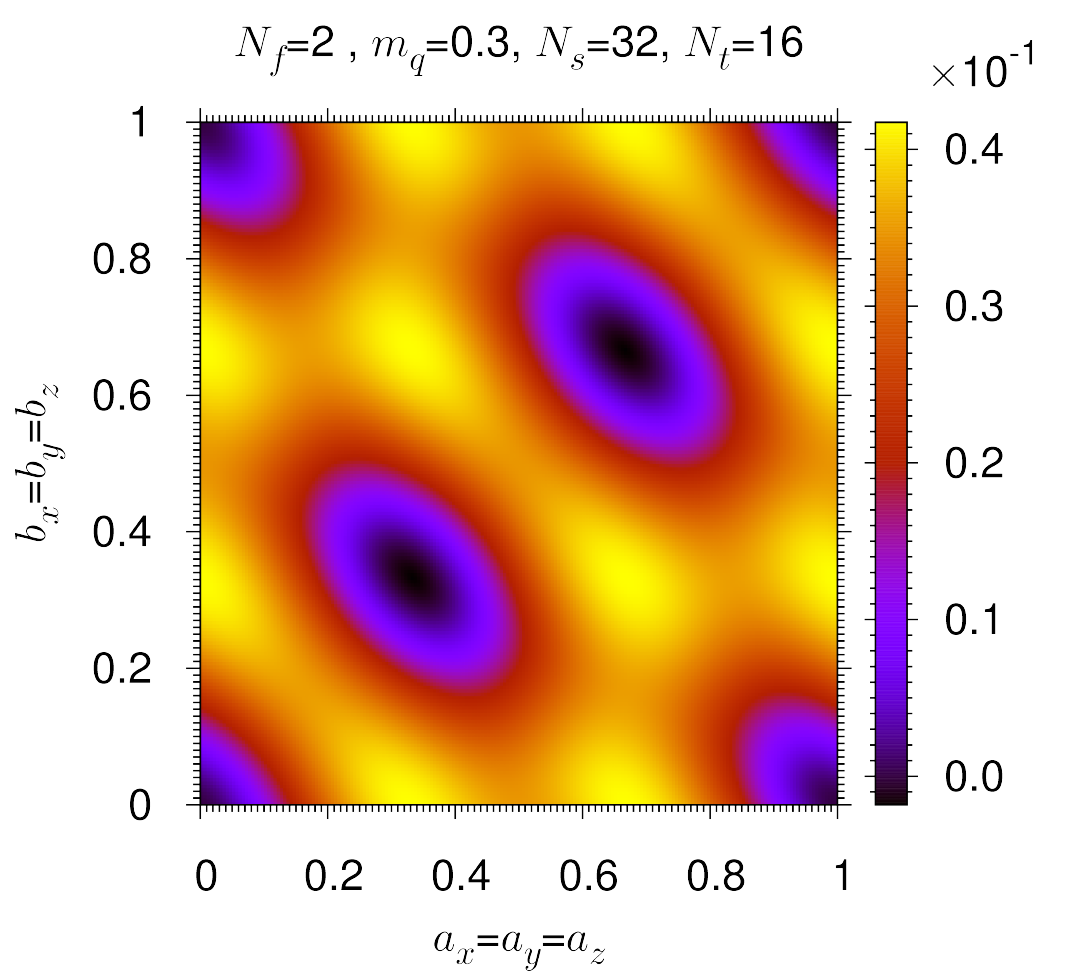}
	\caption{(color online)  The effective potential $V_{\mathrm{eff}}(a, b)$ on $32^3\times 16$ lattice in terms of $a$ and $b$: $m=0.2$ (left) and $m=0.3$ (right).}
	\label{effective potential 2}
\end{figure*}

\section{Analysis at weak coupling}
\subsection{One loop calculation}
Let us now discuss the properties of the phase diagram around the UV fixed point $\beta = \infty$ and $m_q=0$ from analytic computations in perturbation theory.
The free $SU(3)$ gauge theory has the $Z(3)$ center symmetry, but with matter fermions in the fundamental representation, it is broken when the interaction is switched on (i.e. $\beta \neq \infty.$)

The idea is to identify the phase structure in relation to the thermal expectation values of the spatial Polyakov loops. 
For this purpose, we first compute the thermal expectation values of the internal energy (i.e. in the Stefan-Boltzmann law) with a fixed value of the spatial Polyakov loops in the one-loop approximation.
In Ref.\cite{PR89} we discussed the effective potential for the spatial Polyakov loops at zero temperature in detail.
Our computation here is a natural extension to the finite temperature situation.

In the perturbative QCD with the finite spatial volume, the classical zero energy gauge configurations are characterized by flat connections. In the case of our torus lattice, the flat connections are given by the Polyakov loop in each $x,y,z$ directions (in fundamental representation of $SU(3)$):
$$ U_i=  \mathrm{diag} (e^{i2\pi a_i}, e^{ i2\pi b_i}, e^{i 2\pi c_i}) $$ 
with $a_i + b_i + c_i \in \mathbb{Z} $ for $(i=x,y,z)$ from the unitary condition.
Note that $a_i = b_i = c_i = \frac{1}{3}, \frac{2}{3}$ gives a non-trivial center of the gauge group.

As is well-known, the one-loop thermal free energy with $Z(3)$
preserving boundary conditions shows an unphysical infrared divergence
on a finite lattice at finite temperature \cite{Engels:1981ab}. To
avoid this technical problem, we instead compute the thermal
expectation values of the internal energy (which is proportional to
the free energy in conformal field theories after subtracting the
vacuum cosmological constant).
We use the Wilson gauge action for the perturbative computation of the internal energy as in Ref.\cite{PR89}
to circumvent a subtle thermodynamic interpretation of complex energy
poles in the RG improved gauge action at finite $N_t$ \cite{Luscher:1984is}. 
The calculation is then similar to those given in  Appendix D of Ref.\cite{PR89}. We present the details in Appendix
A.

The one-loop thermal expectation values of internal energy including
both fermion loops and gauge field loops
for $N_f=2$ case with  $m_q=0.0, 0.1, 0.2, 0.3, 0.4 \cdots $,
are calculated in the 6 parameter space; $a_i$, $b_i$ in the $x$, $y$
and $z$ directions
on lattices
$16^3\times N_t$, $32^3\times N_t$, $64^3\times N_t$ with $N_t = (4),
16, 32, 64$.
As for four representative cases, we show in Figs.\ref{effective potential 1} and \ref{effective potential 2} the contour map in terms of two parameters; $a$, $b$ putting $a_x=a_y=a_z$; $b_x=b_y=b_z$ for 6 parameters for $32^3\times16$ lattice with $m_q=0.0, 0.1, 0.2$ and $0.3$.

We note the following three points:
\begin{enumerate}
\item
The extrema of the thermal expectation values of the internal energy are given where the spatial Polyakov loops in each directions take an element of the center $Z(3)$ of the gauge group $SU(3)$. There are $4$ species of configurations, in total $3^3$ configurations.
We have the lowest energy when all the Polyakov loops in the spatial directions take non-trivial twisted $Z(3)$ values, $\exp{(\pm i 2/3\pi)}.$
When the spatial Polyakov loop takes a trivial one $\exp(i 0 \pi)=1$, the energy becomes higher. The energy increases as the 
number of trivial directions increase.

\item
When  $m_q \le 3.0/N_t $, the thermal expectation values of the energy 
shows an instability where all of three take the trivial one.  


\item In the thermodynamic limit, the thermal expectation values of the spatial Polyakov loop should be determined by the minimum of the free energy rather than the internal energy. The unphysical divergence of the free energy should be removed in the thermodynamic limit because it is not proportional to the volume. We note that at finite lattice, the one-loop free energy (including the divergence) always takes its minimum value with the $Z(3)$ twisted  boundary condition with $\exp{(\pm i 2/3\pi)}$.
In particular, the free energy difference between the $Z(3)$ trivial boundary condition and $Z(3)$ twisted one is finite irrespective of the divergence because it only comes from the fermion contribution. 

\end{enumerate}

We would like to stress that the above two points $1.$ and $2.$ hold for all the cases we investigate, including the four cases in Figs.\ref{effective potential 1} and  \ref{effective potential 2}. We observe that the thermal expectation values of internal energy is minimized by the $Z(3)$ twisted configurations, and the trivial configuration shows an instability when $m_q=0.0$ but as we increase $m_q$ gradually it becomes metastable around $m_q = 0.2$  (i.e. $m_q \cdot N_t \sim 3.2$): The figures show that the purple parts appear at right bottom and left top corners, which indicates the metastability.

We might suspect that the instability would not appear in the free energy computation (with one-loop perturbative divergence) that should determine the mean thermal average, but we will see below numerical simulations which clearly show the instability in the thermal history (as a mode average).

\begin{figure*}[htb]
\includegraphics[width=7.5cm]{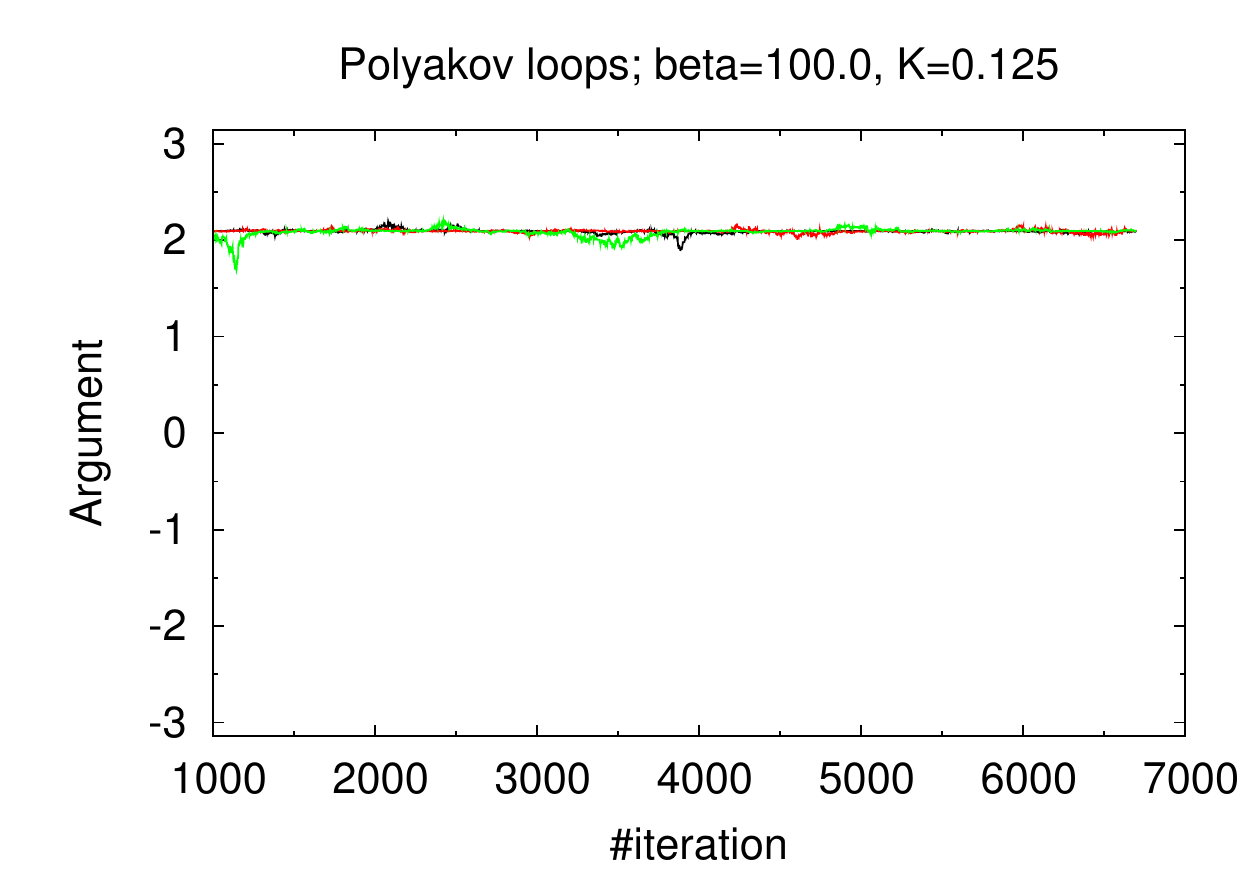}
\hspace{1cm}
\includegraphics[width=7.5cm]{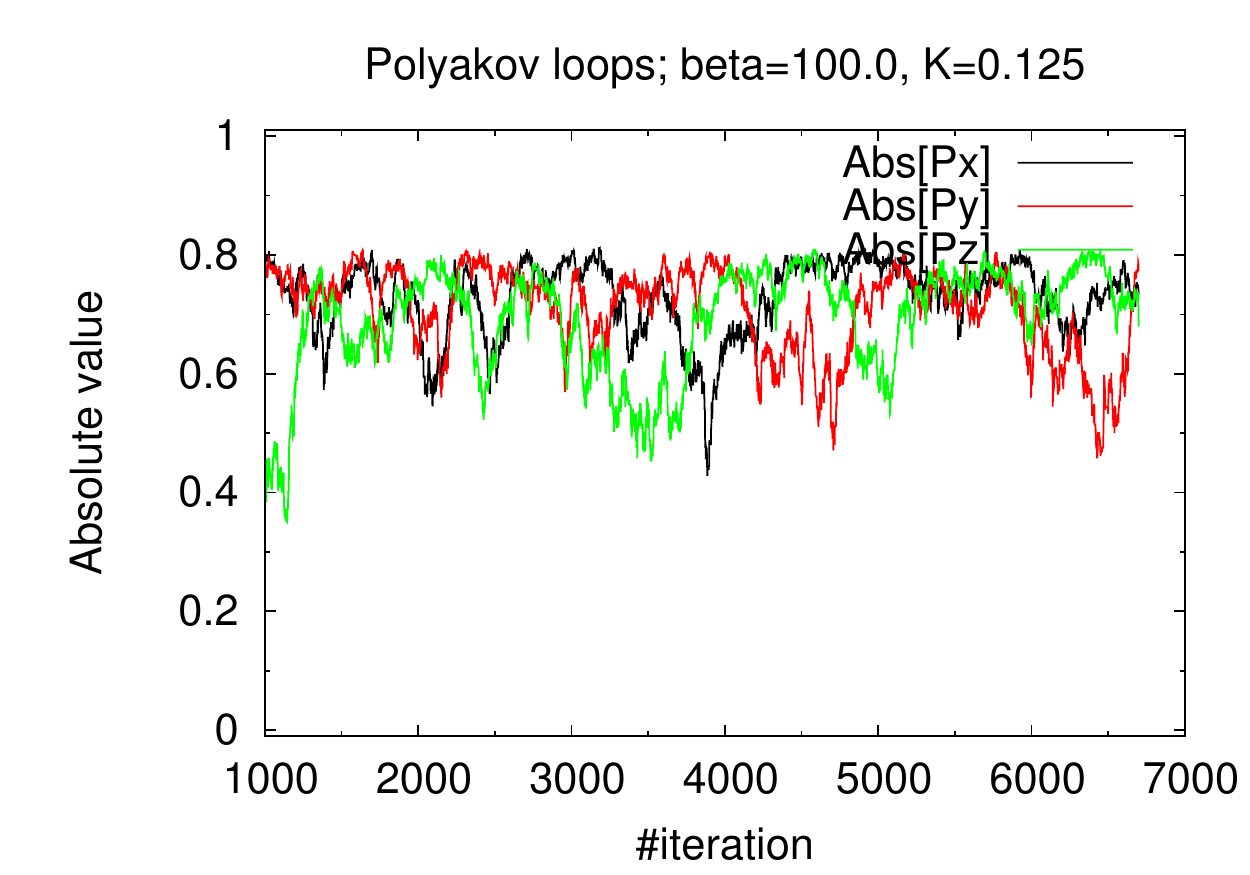}
\caption{(color online)  Time history of the argument and the magnitude value of the spatial Polyakov loops for $\beta=100.0$ and $K=0.125 $ on the $32^3\times 16$ lattice.}
\label{polyakov_beta100}
\end{figure*}

\begin{figure*}[htb]
\includegraphics[width=7.5cm]{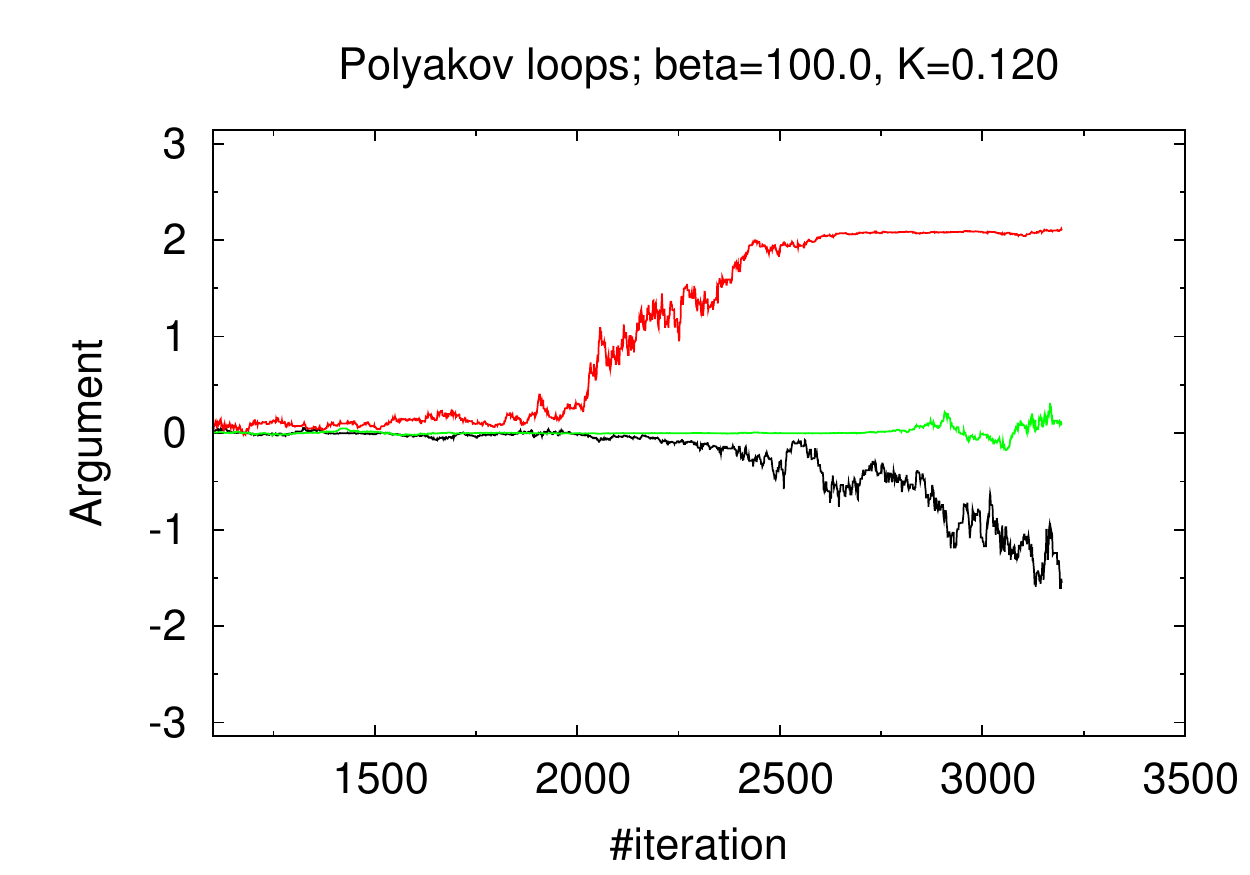}
\hspace{1cm}
\includegraphics[width=7.5cm]{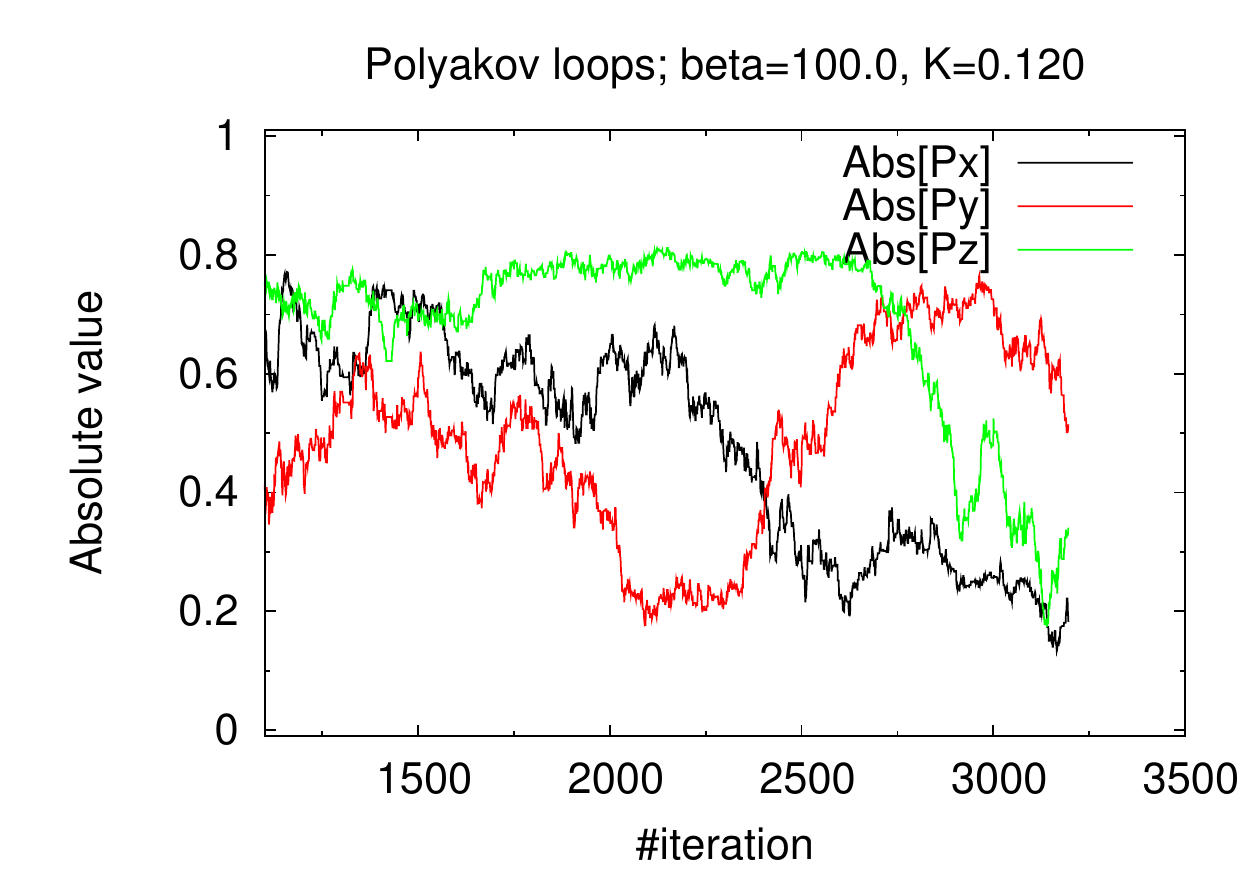}
\caption{(color online)  Time history of the argument and the magnitude value of the spatial Polyakov loops for $\beta=100.0$ and $K=0.120 $ on the $32^3\times 16$ lattice.}
\label{polyakov_k120}
\end{figure*}


\subsection{Numerical results for spatial Polyakov loops}
To compare our analytical results in perturbation theory with the lattice QCD simulations, let us show  in Fig.\ref{polyakov_beta100}
the thermal histories of the magnitude and the angle of the spatial Polyakov loops at $\beta=100.0, K=0.125.$
We see that the angles of all three spatial Polyakov loops clearly stay at $ 2\pi/3=2.09$ (accidentally all three take the plus sign). However the magnitudes are around $0.8.$ The fluctuations are rather large because the lattice size is small.
Thus, at extremely high temperature, the thermal expectation values of physical observables are dominated by the configurations close to the ones with the $Z(3)$ twisted boundary conditions.
In the limit $\beta \to \infty$ the magnitude will approach 1.0. It means that the thermal states obtained at one-loop approximation is a valid description of the numerical simulation.

In Ref.\cite{PR89}, we emphasized that 
 the state where  Polyakov loops in one or two of three spatial directions take unity $1$ instead of $\exp{(\pm i 2/3\pi)}$ is meta-stable at zero temperature 
 and stays for a long simulation time. 
 To choose the true vacuum, we had to (1) first  make simulations at small $\beta$ for long enough time to make transitions among various vacua, and (2) gradually increase the $\beta$.
 
On the other hand we were not able to generate the $Z(3)$ trivial states dynamically at $\beta=100.0$ and $K=0.125$: even if we prepared an ordered state as an initial state, we reach the state where at least one of three spatial Polyakov loops takes $\exp{(\pm i 2/3\pi)}$.
This is consistent with the fact that the $Z(3)$ trivial states are unstable. 

In our finite temperature numerical simulations here, we again encounter the similar behaviors. 
We have made several runs to find the boundary of the local stability of the thermal history. 
Preparing an ordered initial state with heavy quarks and gradually decrease the quark mass, we find that the thermal history of the spatial Polyakov loops becomes unstable around $K=0.120$.
Note that $K=0.120$ corresponds to $m_q=0.18$ (i.e. $m_q \cdot  N_t\simeq 2.9$). We show an example of spatial Polyakov loops at $K=0.120$, Fig.~\ref{polyakov_k120}, where we see that the thermal ensemble is dominated by 
the $Z(3)$ trivial gauge configurations  
about one thousand trajectories. However, it finally collapses to the $Z(3)$ twisted configurations.  In other test cases with $K \geq 0.121$ it decays more quickly.

In order to see the universality of the phase structure at high temperature,
we have also checked that thermal history of spatial Polyakov loops at high temperature, $\beta=100.0$ and $K=0.125,$ with the Wilson gauge action. 
With the Wilson gauge action, the angles of all three spatial Polyakov loops take $\pm 2\pi/3$, while the magnitudes are around $0.7$, which is is a little smaller than $0.8$ with the RG improved action. This suggests that the qualitative picture of the phase diagram presented in this section should be valid in any gauge action as expected from the universality argument.



\subsection{Temporal propagators at weak coupling}
Now let us move onto the study of the temporal propagator in the PS channel.
We are going to compare the free thermal propagator with the $Z(3)$ twisted boundary conditions with the numerical lattice QCD simulations at finite temperature.

\begin{table}
\caption{Effective masses of the temporal PS meson propagator at $\beta=100.0$ (left);
the case of free quarks with the twisted $Z(3)$ boundary condition (center);
the case of free quarks with the trivial boundary condition (right).}
\begin{tabular}{llll}
\hline
\hline
$t$ \, \, \,& simulation \, & twisted\,  & trivial \, \\
\hline
0 & 1.9872(2) & 2.2327 & 2.2329\\
1 & 1.7258(3) & 1.7241 & 1.7247\\
2 & 1.2528(6) & 1.2522 & 1.2516\\
3 & 0.912(1)   & 0.9126 & 0.9065\\
4 & 0.694(1)   & 0.6950 & 0.6794\\
5 & 0.568(1)   &  0.5694 & 0.5459\\
6 &0.507(1)   &  0.5084 & 0.4802\\
7 & 0.483(1)   & 0.4845 &  0.4539\\
\hline
\end{tabular}
\end{table}

\begin{figure}[htb]
\includegraphics[width=7.5cm]{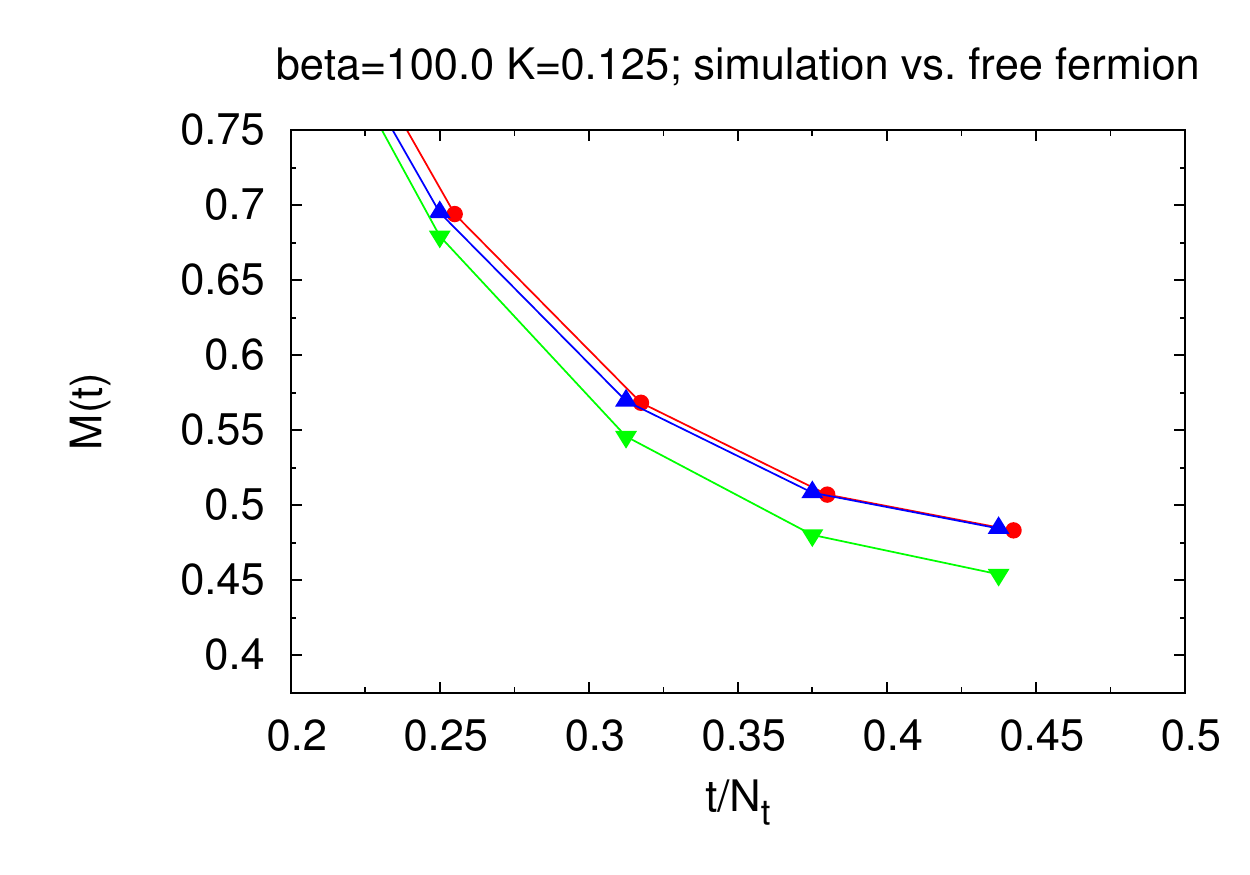}
\caption{ The effective masses at $\beta =100.0$ and K=0.125 (circle; red). The  effective masses of PS meson with the free quarks in the Z(3) twisted boundary condition (triangle; blue) and the Z(3) trivial boundary condition (inverted triangle; green). The red points are shifted horizontally right by 0.005 in order that we are able to see the both of  blue points and red points.}
\label{effm_32x16_free_32x16}
\end{figure}

We begin with  the numerical result of effective mass at $\beta=100.0$ and $K=0.125$.
We present the data in Table 2 together with that of the free fermion with the twisted boundary condition. We see that the effective mass is remarkably in agreement with that of the free fermion with the twisted boundary condition as shown both in Table 2 and in Fig.\ref{effm_32x16_free_32x16}.
Since the differences are $\sim 0.2 \%$ for $t=3 - 7$, and they are too small to observe in the figure,
the red points (lattice simulation) are shifted right by 0.005 in order to see the blue points (perturbative computation).
In comparison, if we compute the free thermal propagator with the trivial boundary conditions,
the result is different from the numerical result with order $\sim 6.0\%.$ 

\begin{figure*}[htb]
\includegraphics[width=7.5cm]{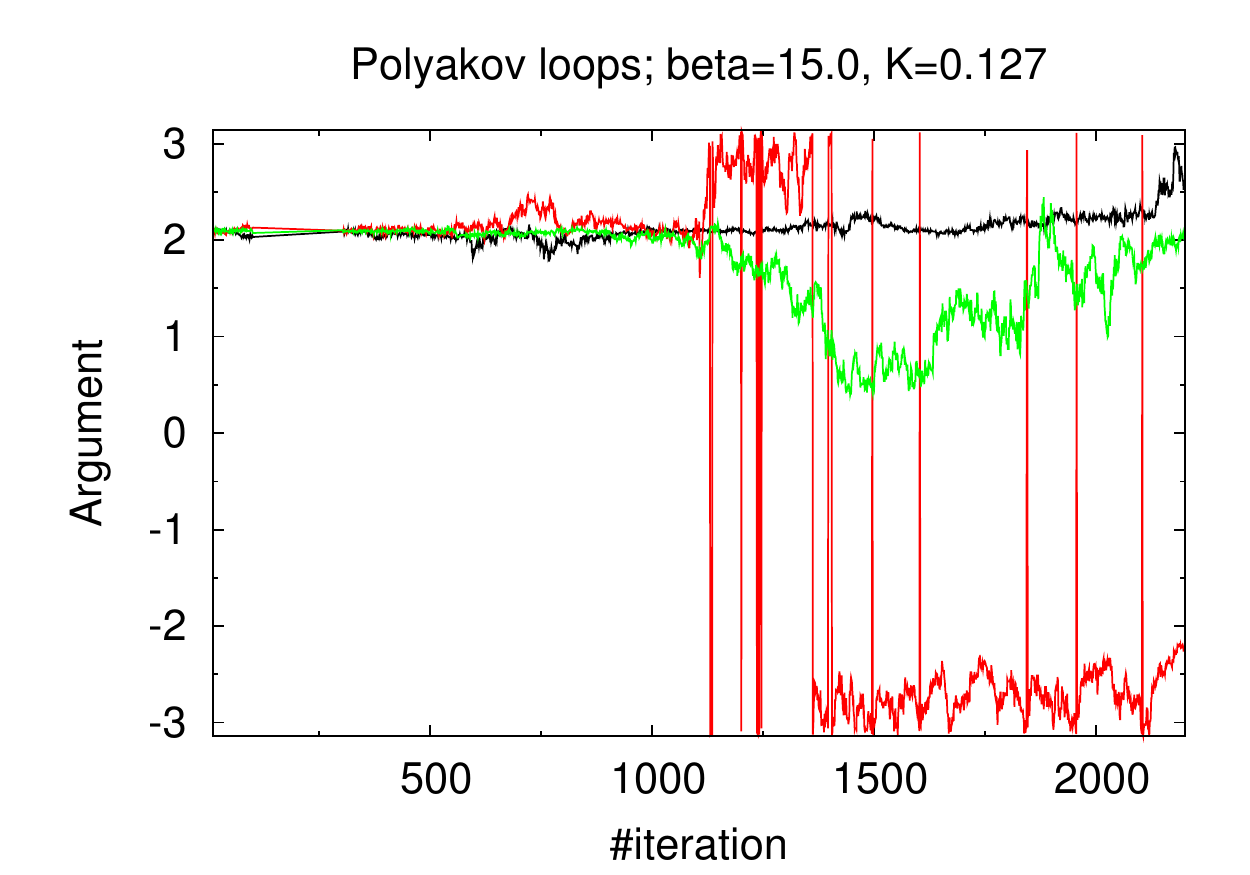}
\hspace{1cm}
\includegraphics[width=7.5cm]{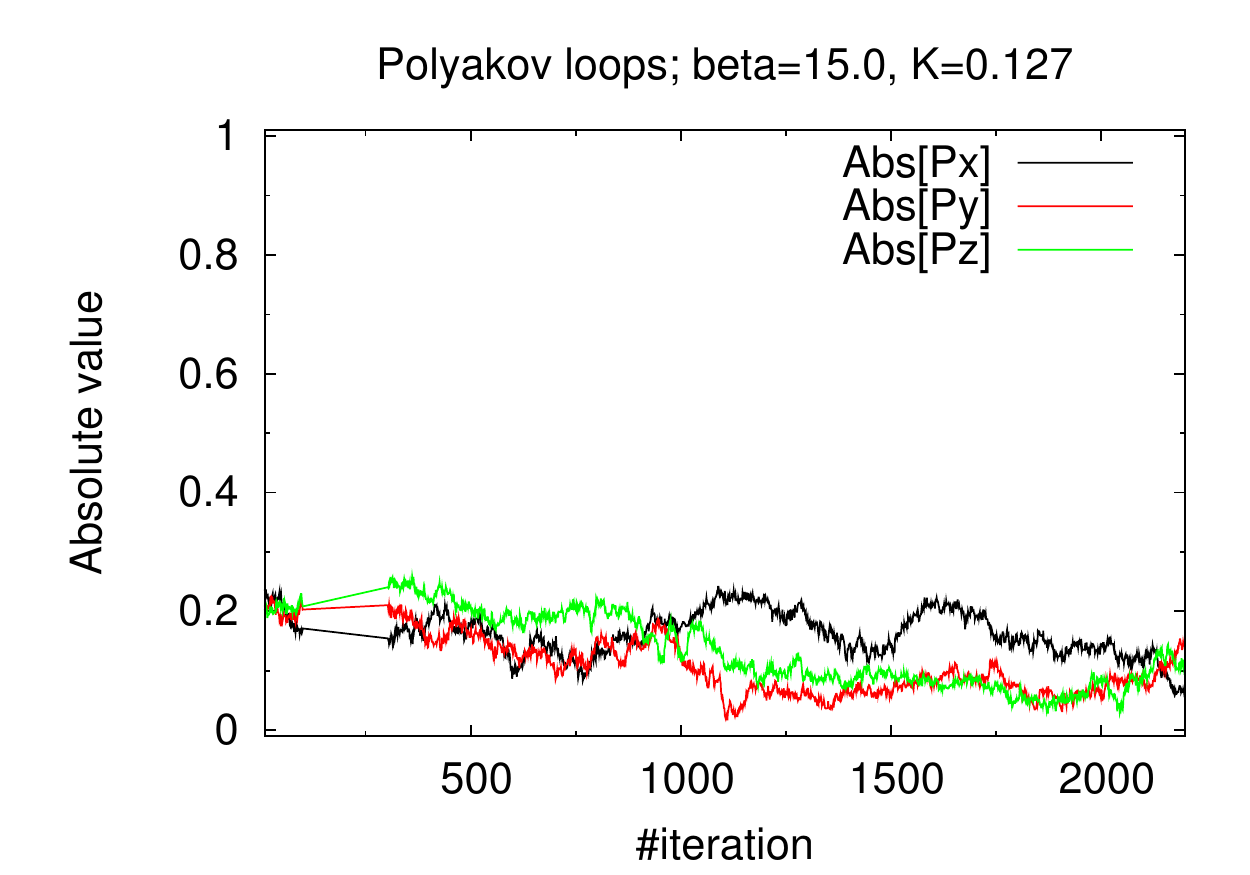}
\caption{(color online)  Time history of the argument and the magnitude value of the spatial Polyakov loops for $\beta=15.0$ and $K=0.127 $ on the $32^3\times 16$ lattice.}
\label{poyakov_beta15}
\end{figure*}

\section{Simulations at strong coupling}
\subsection{Spatial Polyakov loops and propagators along the massless line}
In order to approach the strongly coupled regime, we decrease $\beta$ along the massless quark line toward the chiral transition point $\beta=2.9$; $$\beta=15.0, 10.0, 6.0, 5.0, 4.0, 3.0.$$
As an example, we show the thermal histories of the magnitude and the angle of the spatial Polyakov loops at $\beta=15.0,$ in Fig.
\ref{poyakov_beta15}.
We see that the fluctuations become larger as $\beta$ decreases, and the transition between different Z(3) center configurations occurs more frequently.

\begin{figure*}[htb]
\includegraphics[width=7.5cm]{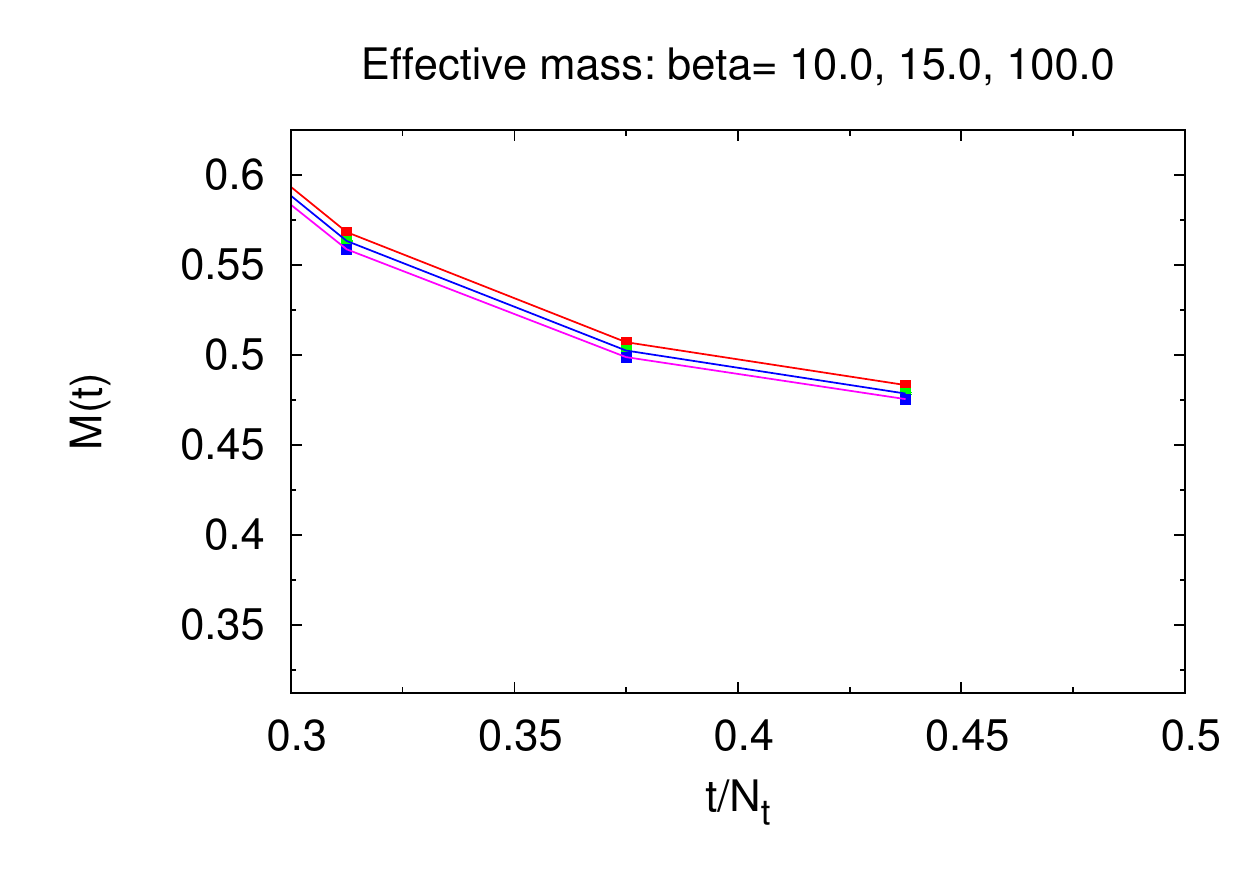}
\hspace{1cm}
\includegraphics[width=7.5cm]{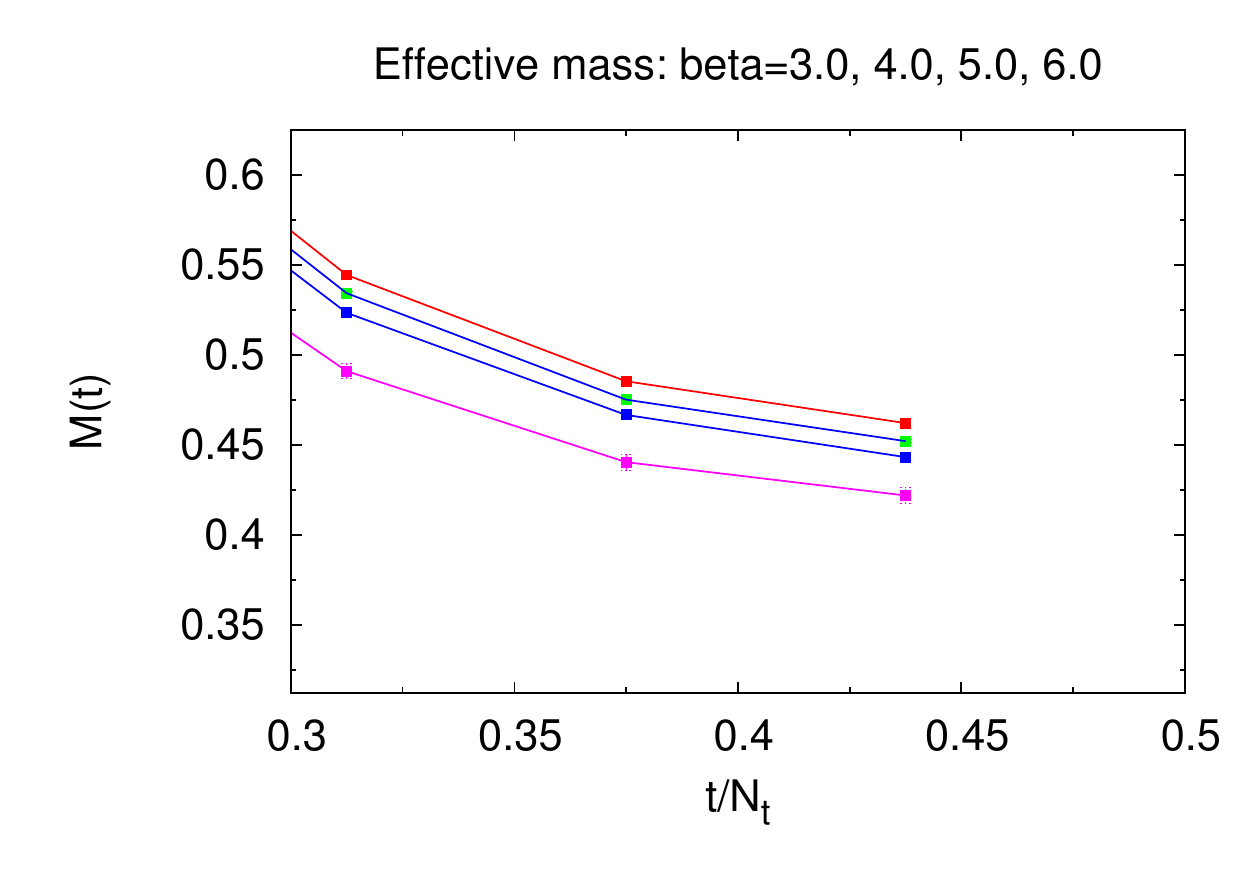}
\caption{(color online) The effective mass for $\beta=100.0, 15.0, 10.0$ in order from the top to the bottom
(left) and $\beta=6,0, 5.0, 4.0, 3.0$ in order from the top to the bottom (right).}
\label{beta30_100}
\end{figure*}

Although the spatial Polyakov loops behave more differently with more transitions as we decrease $\beta$,
the effective mass for $\beta= 15.0, 10.0$ and $\beta=100.0$
are identical within $1\%$ difference for $t\geq 3$. See Table 3.
Overlaying the three data we are not able to distinguish them as shown in Fig.\ref{beta30_100}.
We interpret that the thermal states with $\beta=100.0$ down to $\beta=10.0$
are still dominated by the $Z(3)$ twisted gauge configurations and the quarks behave very similarly to the free massless fermions with the $Z(3)$ twisted boundary condition.

When we further decrease $\beta$ as $\beta=6.0, 5.0, 4.0$, the effective mass at large distance slightly deviates from the curve at $\beta=100.0$. See Table 4.
When we reach $\beta=3.0$ the effect mass clearly deviates from it as can be seen in Fig.\ref{beta30_100}. 
This deviation at $\beta=3.0$ is expected, since  $\beta=2.9$ is the chiral transition point and at that point we are no longer in the weakly coupled regime.

\begin{table}
\caption{Effective masses of the temporal PS meson propagators at $\beta=100.0. 15.0$ and $10.0$.}
\begin{tabular}{llll}
\hline
\hline
$t$ \, \, \, \,& $\beta=100.0$ \, &   $\beta=15.0$ \, & $\beta=10.0$ \\
\hline
0 & 1.98727(1) & 1.95298(9) & 1.8819(2)\\
1 & 1.72584(2) & 1.7096(2) & 1.6775(3)\\
2 & 1.25283(4) & 1.2452(2) & 1.2284(3)\\
3 &  0.91225(7) & 0.9068(3) & 0.8977(3)\\
4 &  0.69417(5) & 0.6890(3) & 0.6827(3)\\
5 &  0.56837(9) & 0.5635(4) & 0.5589(4)\\
6 &  0.50725(10) & 0.5026(4) & 0.4989(4)\\        
7 &  0.48338(10) & 0.4787(5) & 0.4756(4)\\
\hline
\end{tabular}
\end{table}

\begin{table}
\caption{Effective masses of the temporal PS meson propagators at $\beta=6.0, 5.0, 4.0$ and $3.0$.}
\begin{tabular}{lllll}
\hline
\hline
$t$ \, \, \, \,& $\beta=6.0$ \, &   $\beta=5.0$ \, & $\beta=4.0$\, & $\beta=3.0$ \\
\hline
0 & 1.8260(4) & 1.7859(3)  & 1.7400(14) & 1.6167(16)\\
1 & 1.6480(8) & 1.6275(7) & 1.5998(15) & 1:5212(29)\\
2 & 1.2093(8) & 1.196(10)	 & 1.1750(22) & 0.7994(53)\\
3 & 0.8824(9) & 0.8714(9) &  0.8525(21) &  0.7310(53)\\
4 & 0.6675(9) & 0.6568(8) & 0.6414(28) &  0.5983(45)\\
5 & 0.5445(10) & 0.5344(8) &  0.5235(25) &  0.4911(40)\\
6 & 0.4855(11) & 0.4753(8) &  0.4668(17) & 0.4406(43)\\
7 & 0.4623(11) & 0.4522(8) &  0.4434(16)  & 0.4220(43)\\
\hline
\end{tabular}
\end{table}

\subsection{Spatial Polyakov loops and temporal propagators with massive quarks}
Now let us discuss the case with massive quarks at finite $\beta$.
As long as the quark mass is smaller than the critical mass, the spatial Polyakov loops still take a non-trivial $Z(3)$ element. On the other hand, when the quark mass becomes heavier, it  takes the trivial $Z(3)$ element.\footnote{Although in the small $K$ expansion, the Z(3) twisted configuration is slightly favored\cite{finite size}, the Z(3) trivial configuration is chosen  after thermalization with ordered initial state}
At a certain critical quark mass, we find these two states co-exist as in \cite{PR87}\cite{PR89}.

\begin{figure*}[htb]
\includegraphics[width=7.5cm]{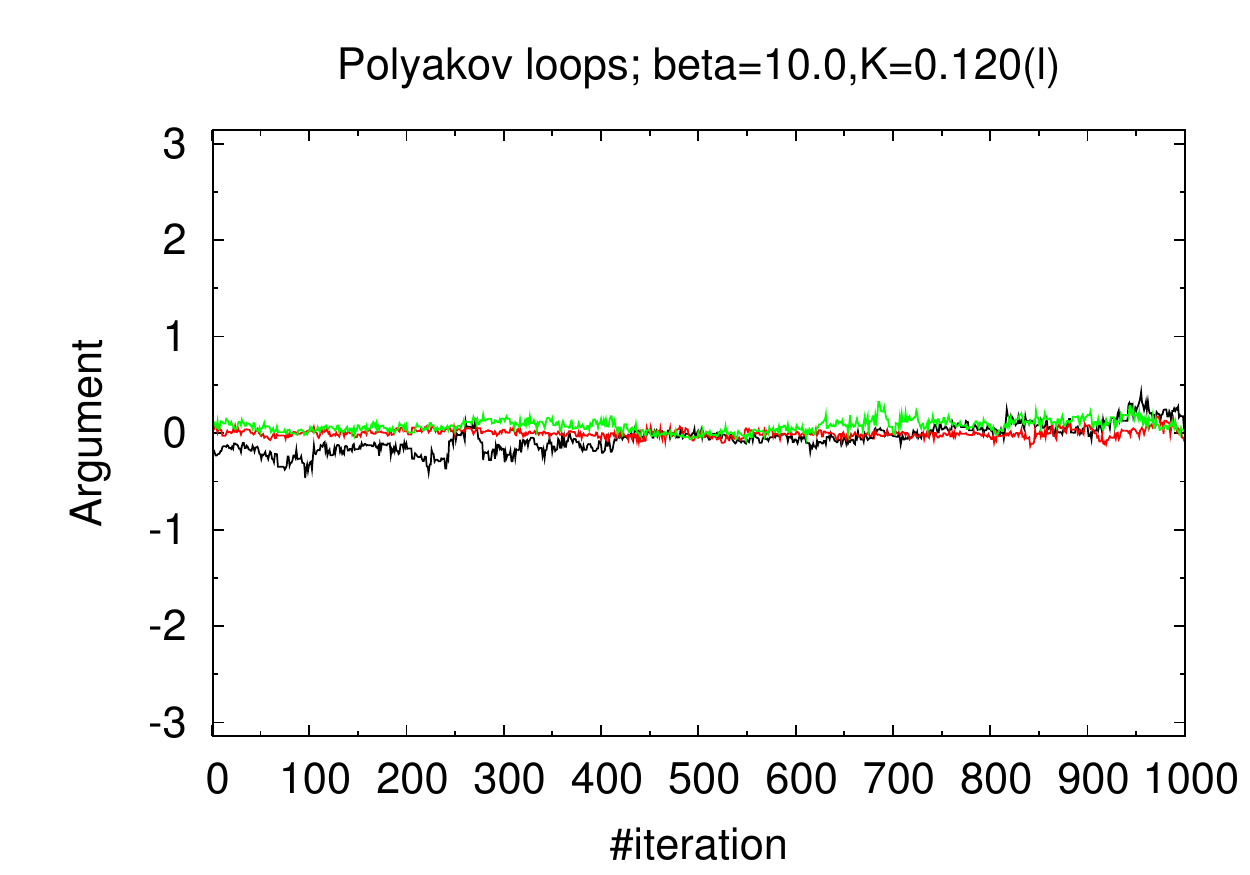}
\hspace{1cm}
\includegraphics[width=7.5cm]{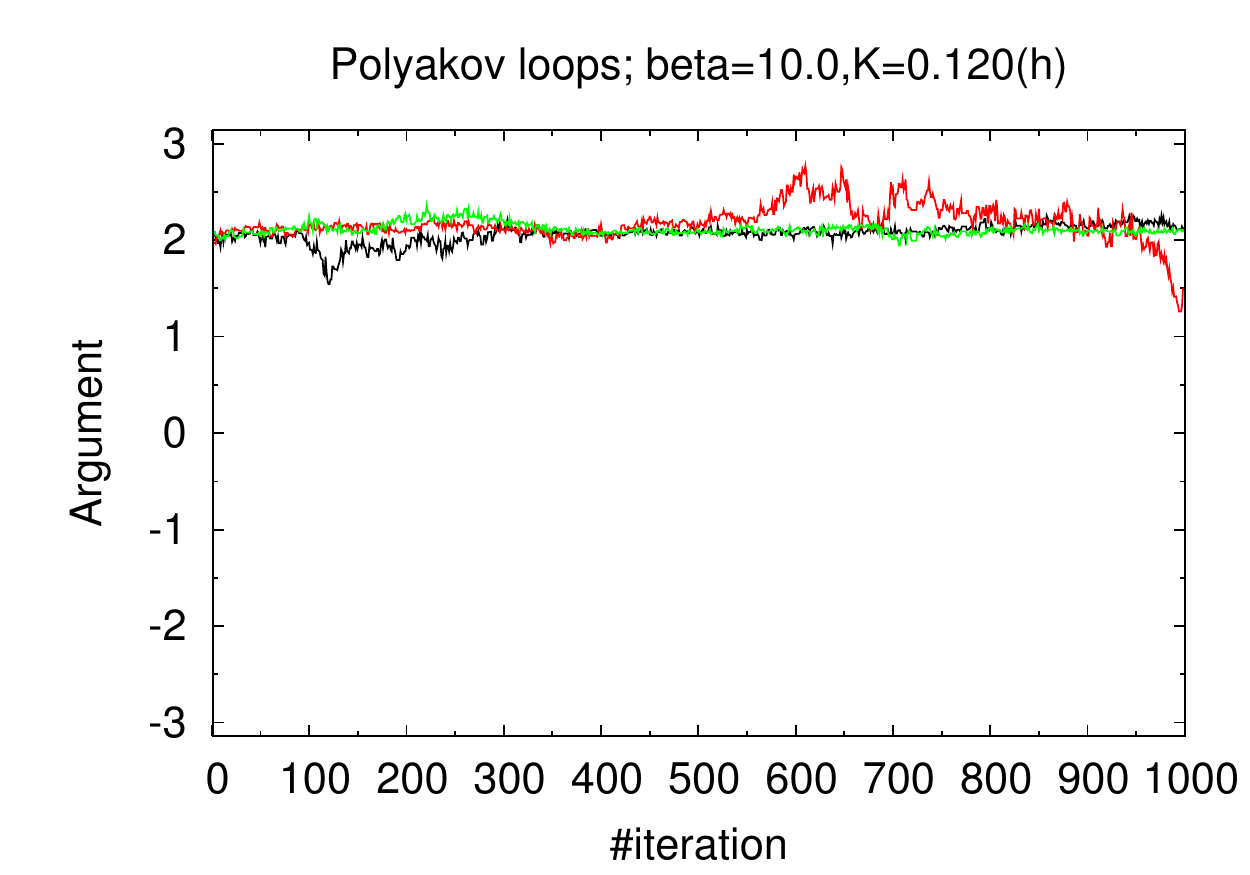}
\caption{(color online) Time history of the argument of the Polyakov loops for $\beta=10.0$ and $K=0.120 $ on the $32^3\times 16$ lattice: (left panel) the cold start: (right panel) the hot start.}
\label{beta10k120}
\end{figure*}

As a demonstration, we elaborate on the transition at $\beta=10.0$.
In Fig.\ref{beta10k120} we show the histories of angles of the spatial Polyakov loops at $K=0.120.$
The initial configuration in the right panel is that at $K=0.125$ (the hot start), while in the left panel it is that of $K=0.110$ (the cold start).
We see clear difference between the two. The spatial Polyakov loops take the twisted $Z(3)$ value in the right panel, while the trivial one in the left panel.

To further investigate the existence of the transition, we, in turn, study hadronic quantities at $\beta=10.0$ and $K=0.120$. One interesting comparison is the PS channel effective mass at $t=7$.
With the hot start we obtain $m_q=0.301(1)$ and $m_{\mathrm{PS}}=0.8198(2)$, while with the cold start we have $m_q=0.301(1)$ and $m_{\mathrm{PS}}=0.8188(4)$. 
It is expected that the quark masses $m_q$ are identical, but given the change of the spatial Polyakov loop expectation values, it may be a little surprising that the effective masses appear almost identical. However, we claim that the difference is non-zero $\delta m_{\mathrm{PS}}=0.0010(6)$ with a controlled error estimate.

To see this, we note that in the free one-loop computation with $m_q=0.3$  we have $m_{\mathrm{PS}}=0.792643$ with the Z(3) twisted boundary condition, while we have $ m_{\mathrm{PS}}=0.790817$ with the Z(3) trivial boundary condition: the difference $\delta m_{\mathrm{PS}}^{\mathrm{free}} = 0.00182$ is the same order as we observed in the numerical simulation above. In our earlier studies on the  $16^3\times 64$ lattice \cite{Ishikawa:2013iia}, the transition was visibly seen in the effective mass difference $\delta m_{\mathrm{PS}} \sim 0.4 $ in good agreement between one-loop computation and numerical simulations, suggesting that jump in the temporal propagator indicates the transition between region in which the $Z(3)$ trivial gauge configuration dominates and the one in which the non-trivial gauge configuration does. We think that the much smaller but non-zero difference here can be attributed to the similar transition.

	

\section{Discussions}
In this article we have proposed  a new perspective on quarks and gluons at high temperature.
The quarks and gluons at very high temperature are free but 
effectively described by the $Z(3)$ twisted boundary condition.
As long as the quark mass is sufficiently small, this picture is valid, and the situation is similar to the many flavor conformal QCD.


Let us discuss the transition between the small quark mass region and the large quark mass region.
The  end point of the red curve at $\beta=\infty$ shown in Fig.1 should correspond to the boundary for the stability of the $Z(3)$ trivial configurations.
We estimate $m_q \simeq 3.0 \,T$ from the one loop computation of internal energy with fixed boundary conditions.


The red line runs toward the chiral phase transition point.
Since there is no order parameter of $Z(3)$ symmetry except $\beta =\infty$, 
it is not possible to identify the red line at small $\beta$ in a strict sense.
However we have shown that we are able to identify the red line from the behavior of the spatial Polyakov loops and a small jump in temporal propagators.



The results of this work are based on the simulation of a fixed
lattice size, and it remains an important issue to address the
thermodynamic limit and the continuum limit more systematically. We expect
that the effects of the spatial boundary condition will become less
dominant in the thermodynamic limit, but it would be interesting to
directly see how different choices of the boundary condition (e.g.
anti-periodic spatial boundary condition for fermions) affect our
discussions, in particular, in the thermodynamic limit.
In our recent work Ref. \cite{PL}, we have developed a new RG method
to locate the renormalization group fixed point at zero temperature by
comparing simulations with different lattice sizes.
We believe that the similar method at finite temperature will help us
determine the phase structure of finite temperature QCD more precisely
and clarify the nature of the thermodynamic limit. We are going to
report the progress in the near future.





\section*{Acknowledgement}
We would like to thank Akira Ukawa for reading through the manuscript, and Kazuyuki Kanaya and Tetsuo Hatsuda for useful discussion.

The calculations were performed on 
Hitachi SR16000 at KEK under its Large-Scale Simulation Program
and HA-PACS computer at CCS, University of Tsukuba
under HA-PACS Project for advanced interdisciplinary computational sciences by exa-scale computing technology.

The work by Yu.~Nakayama is supported by the World Premier International Research Center Initiative (WPI Initiative), MEXT, Japan.\\


\appendix

\section{Internal energy in one-loop approximation}

In this appendix we compute the internal energy $U$ in the one-loop approximation for the Wilson gauge and Wilson fermion actions. In the canonical ensemble, the internal energy is defined by
\begin{align}
    U = \dfrac{\partial}{\partial T^{-1}} (F(T)T^{-1}),
\end{align}
where $T$ is the temperature and $F(T)$ is the Helmholtz free energy. 
In the main text, we have explained the reason why 
we calculate the internal energy instead of the free energy.


On the finite lattice, the bosonic internal energy density for a free field theory with a fixed momentum $\bm{p}$ is given by (in the unit $a=1$)
\begin{align}
u_B(\bm{p}) &=
\left[
\dfrac{1}{2} E^{\mathrm{lat}}_B(\bm{p})
+ \dfrac{E_B^{\mathrm{lat}}(\bm{p})}{e^{N_t E_B^{\mathrm{lat}}(\bm{p})}-1}
\right] \cr
&+(\mbox{terms independent of $\bm{p}$}).
\end{align}
Similarly, the fermionic internal energy density is
\begin{align}
&        u_F(\bm{p}) \cr
&= -
\left[
\dfrac{1}{2}E^{\mathrm{lat}}_F(\bm{p})
+\dfrac{1}{2}\log(M(\bm{p})+1)
- \dfrac{E_F^{\mathrm{lat}}(\bm{p})}{e^{N_t E_F^{\mathrm{lat}}(\bm{p})}+1}
\right] \cr
& +(\mbox{terms independent of $\bm{p}$}).
\end{align}
Here the on-shell energy for the Wilson gauge action, $E_B^{\mathrm{lat}}$ is
\begin{align}
E_B^{\mathrm{lat}}(\bm{p}) &=2\sinh^{-1}|\tilde{\bm{p}}|, \cr
|\tilde{\bm{p}}|^2 &= 
4\sin^2\left(\dfrac{p_x}{2}\right)
+4\sin^2\left(\dfrac{p_y}{2}\right)
+4\sin^2\left(\dfrac{p_z}{2}\right) , 
\end{align}
and the on-shell energy for the Wilson fermion, $E_F^{\mathrm{lat}}$ is
\begin{align}
E^{\mathrm{lat}}_F(\bm{p}) &=
\cosh^{-1}\left[
\dfrac{
	|\mathring{\bm{p}}|^2 + (M(\bm{p})+1)^2+1}{2(M(\bm{p})+1)}
\right],\cr
|\mathring{\bm{p}}|^2 &= \sin^2(p_x)+\sin^2(p_y)+\sin^2(p_z),\cr
M(\bm{p}) &= m_f + 3
-\cos\left(p_{x}\right)
-\cos\left(p_{y}\right)
-\cos\left(p_{z}\right).
\end{align}

As discussed in section 4.1, in the perturbative limit of the QCD, we parametrize a flat connection on the three spatial torus lattice as
\begin{align}
  P_j=\mathrm{diag}(e^{i2 \pi a_j},e^{i2\pi b_j},e^{i2\pi c_j})
\end{align}
%
with $a_i + b_i + c_i \in \mathbb{Z} $ with $(i=x,y,z)$ from the unitary condition.
We will denote the vector  $(a_i ,b_i , c_i)$  with  $(i=x,y,z)$ as $(\bm{a},\bm{b},\bm{c})$ in the following.
 Then the internal energy is a function of the Polyakov loops $U(\bm{a},\bm{b},\bm{c}).$
The total internal energy in $\mathbb{T}^3$
with a non-trivial background gauge field is evaluated as
    \begin{align}
\delta U(\bm{a},\bm{b},\bm{c}) &=
U(\bm{a},\bm{b},\bm{c}) - U(\bm{0},\bm{0},\bm{0}).
    \end{align}
$U$ involves all contributions from the momentum shift
after the singular gauge transformation;
    \begin{align}
        U(\bm{a},\bm{b},\bm{c}) &=
2 \left[
\sum_{c=1}^{8}
  \sum_{n_j=A^{(c)}_j}^{A^{(c)}_j+N_s-1} u_B(2\pi \bm{n}/N_s)
\right]
\notag\\
&
+ 4 N_f \left[ 
\sum_{c=1}^{3}
  \sum_{n_j=a^{(c)}_j}^{a^{(c)}_j+N_s-1} u_F(2\pi \bm{n}/N_s)
\right]\cr
&+(\mbox{terms independent of $(\bm{a},\bm{b},\bm{c})$}),
    \end{align}
where the momentum sum is taken as 
$n_x=a_x, a_x + 1, \cdots, a_x + N_s -1$, etc. 
The momentum shift is explicitly given by
\begin{align}
\bm{A}^{(1)} &= \bm{a}-\bm{b},\quad&
\bm{A}^{(2)} &= \bm{b}-\bm{a},\quad&
\bm{A}^{(3)} &= \bm{c}-\bm{a},\notag\\
\bm{A}^{(4)} &= \bm{a}-\bm{c},\quad&
\bm{A}^{(5)} &= \bm{b}-\bm{c},\quad&
\bm{A}^{(6)} &= \bm{c}-\bm{b}, \cr
\bm{A}^{(7)} &= \bm{A}^{(8)} = 0,\notag\\
\bm{a}^{(1)} &= \bm{a},\quad&
\bm{a}^{(2)} &= \bm{b},\quad&
\bm{a}^{(3)} &= \bm{c}.
\end{align}

The resulting potential was evaluated and plotted in the main text.

\end{document}